\documentclass[12pt]{article}
\usepackage{amsmath,amssymb}
\newcommand{\be}{\begin{equation}}
\newcommand{\ee}{\end{equation}}
\newcommand{\bea}{\begin{eqnarray}}
\newcommand{\eea}{\end{eqnarray}}

\newcommand{\p}[1]{(\ref{#1})}
\newcommand{\lb}{\label}

\begin{document}

\begin{center}
{\Large\bf M2-Branes in ${\cal N}=3$ Harmonic Superspace}
\vspace{0.6cm}

{\large\bf
Evgeny Ivanov}
\vspace{0.4cm}

{\it Bogoliubov Laboratory of Theoretical Physics,
JINR, \\
141980, Dubna, Moscow Region, Russia}\\
{\tt eivanov@theor.jinr.ru}\\[8pt]
\end{center}
\vspace{0.6cm}


\begin{center}
{\it Invited Talk at the Conference ``Selected Topics in Mathematical and Particle Physics'', \\In
Honor of the 70-th Birthday of Jiri Niederle, Prague, 5 - 7 May 2009}
\end{center}
\vspace{0.5cm}

\begin{abstract}
\noindent
We give a brief account of the recently proposed ${\cal N}=3$ superfield formulation
of the ${\cal N}=6, 3D$ superconformal theory of Aharony {\it et al} (ABJM) describing a low-energy limit
of the system of multiple M2-branes on the AdS$_4\times S^7/\mathbb{Z}_k$ background. This formulation
is given in harmonic ${\cal N}=3$ superspace and reveals a number of surprising new features.
In particular, the sextic scalar potential of ABJM arises at the on-shell component level as the result of eliminating
appropriate auxiliary fields, while there is no explicit superpotential at the off-shell superfield level.
\end{abstract}

\section{Preliminaries: AdS/CFT}
\subsection{AdS/CFT in type IIB superstring}
As the starting point, I recall the essentials of the original AdS/CFT correspondence
(for details see \cite{AdsCFT} and references therein).

It is the conjecture that type IIB superstring on AdS$_5\times $ S$^5$ is in some sense dual to maximally
supersymmetric ${\cal N}=4, 4D$ super Yang-Mills (SYM) theory.
This hypothesis is to a large extent based upon the coincidence of the symmetry groups of both theories.
Indeed,
$$
AdS_5\times S^5 \sim \frac{SO(2,4)}{SO(1,4)}\times
\frac{SO(6)}{SO(5)} \subset  \frac{SU(2,2|4)}{SO(1,4)\times SO(5)}\,,
$$
so the superisometries of this background constitute the supergroup $SU(2,2|4)$.
On the other hand, the supergroup $SU(2,2|4)$ defines superconformal invariance of
${\cal N}=4$ SYM, with $SO(2,4)$ and $SO(6) \sim
SU(4)$ being, respectively, $4D$ conformal group  and R-symmetry group.
\vspace{0.2cm}

Some related salient features of the AdS/CFT correspondence are as follows.

\begin{itemize}
\item
AdS$_5\times $S$^5$ (plus a constant closed 5-form on $S^5$) is the bosonic
``body'' of the maximally supersymmetric curved solution $\frac{SU(2,2|4)}{SO(1,4)\times SO(5)}$ of
type IIB, $10D$ supergravity. It  preserves {\bf 32} supersymmetries.

\item
${\cal N}=4$ SYM action with the gauge group
$U(N)$ is the low-energy limit of a gauge-fixed action
of a stack of $N$ coincident D3-branes
on AdS$_5\times $S$^5$: ${\bf 4}$ worldvolume co-ordinates of the latter system become the Minkowski space-time co-ordinates,
while ${\bf 6}$ transverse ($u(N)$ algebra-valued) D3-brane co-ordinates
yield just ${\bf 6}$ scalar fields of the nonabelian ${\cal N}=4, 4D$ gauge multiplet.

\item
This system has the following {\it on-shell} content: ${\bf 6}$ bosons and ${\bf 16/2} = {\bf 8}$
fermions (all $u(N)$ algebra valued); ${\bf 2}$ ``missing'' bosonic degrees of freedom which are required by world-volume ${\cal N}=4$
supersymmetry come from a gauge field. This is a ``heuristic'' explanation why just
D3-branes, with the gauge fields contributing non-trivial degrees of freedom on shell, matter in the case of the
AdS$_5$/CFT$_4$ correspondence.

\end{itemize}

\subsection{AdS/CFT in M-theory}
Recently, there has been a surge of interest in another example of AdS/CFT duality,
this time related to M-theory and type IIA superstring.

The fundamental (though not explicitly formulated as yet) M-theory can be defined as a strong-coupling limit of type IIA, $10D$ superstring
with $11D$ supergravity as the low-energy limit. It has the following maximally supersymmetric classical curved
solution:
$$
AdS_4\times S^7 \sim \frac{SO(2,3)}{SO(1,3)}\times
\frac{SO(8)}{SO(7)} \subset  \frac{OSp(8|4)}{SO(1,3)\times SO(7)}
$$
(plus a constant closed 7-form on $S^7$), which preserves {\bf 32} supersymmetries.
\vspace{0.1cm}

When trying to treat this option within the general AdS/CFT correspondence (like the previously discussed  AdS$_5\times$ S$^5$ example),
there arise the following natural questions.
\begin{itemize}
\item What
is the CFT dual to this geometry?
\begin{enumerate}
\item
It should be some 3D analog of ${\cal N}=4$ SYM and should arise as
a low-energy limit of multiple M2-branes (membranes of M-theory, analogs of D3-branes of type IIB superstring).
\item
Hence it should contain {\bf 8} (gauge algebra valued)
scalar fields which originate from the transverse co-ordinates of
M2-branes.
\item
It should contain off-shell {\bf 16} physical fermions
({\bf 16} other fermionic modes can be gauged away by
the relevant $\kappa$ symmetry).
\item
Finally, it should be superconformal, with $OSp(8|4)$ realized as
${\cal N}=8,3D$ superconformal group.
\vspace{0.1cm}
\end{enumerate}

\item On shell there should be ${\bf 8 + 16/2 = 8+8}$
degrees of freedom. Hence the gauge fields should not contribute any
degree of freedom on shell in this special case (in a drastic contrast with the ``type IIB /${\cal N}=4$ SYM'' correspondence).
\end{itemize}
\vspace{0.1cm}

\noindent The unique possibility which meets all these demands is that the dual theory
is some supersymmetric extension of {\it Chern-Simons} gauge theory \cite{Schwarz04}.

\section{Chern-Simons theories}
The standard bosonic Chern-Simons (CS) action is as follows
\bea
&& S_{cs} = \frac{k}{4\pi} \mbox{Tr} \int d^3x \epsilon^{mns}\left(A_m\partial_n A_s
+ \frac{2i}{3} A_m A_nA_s \right) \lb{CS} \\
&& \Rightarrow \quad {\cal F}_{mn} = \partial_m A_n - \partial_n A_m + [A_m, A_n] = 0, \nonumber
\eea
i.e. the YM field $A_n$ is pure gauge on shell.

The ${\cal N}=1$ superextension of the CS action is obtained by extending $A_n$ to ${\cal N}=1$ gauge supermultiplet
\be
A_n \Rightarrow (A_n, \chi^\alpha), \quad \alpha = 1,2\,; \qquad {\cal L}_{cs}(A)
\Rightarrow  {\cal L}_{cs}(A) - \mbox{Tr} (\bar\chi \chi)\,. \lb{SCS}
\ee
The fermionic field $\chi$ is auxiliary, and no dynamical
(Dirac) equation for it appears. The same phenomenon takes place in the case
of ${\cal N}=2$ and ${\cal N}=3$ superextensions of the pure CS action. The physical fermionic fields (having standard kinetic terms) can
appear only from the matter supermultiplets coupled to the CS one.

Keeping in mind these general properties of supersymmetric Chern-Simons theories, Schwarz assumed \cite{Schwarz04} that the theory dual to
AdS$_4\times$S$^7$ must be ${\cal N}=8$
superextension of the $3D$ CS theory, i.e. one should deal with the
on-shell supermultiplet $(A_m, \phi^I, \psi^B_\alpha), I = 1,... 8, \quad B =
1, ... 8\,$.

How to gain physical kinetic terms for {\bf 16} ($u(N)$ algebra-valued) fermions? The recipe: place the latter
into matter multiplets of the manifest ${\cal N}=1$,
${\cal N}=2$ or ${\cal N}=3$ supersymmetries, consider the relevant combined ``CS + matter'' actions  and realize
extra supersymmetries as the hidden ones mixing the CS supermultiplet with the matter multiplets.

\setcounter{equation}{0}

\section{BLG and ABJM models}
 \subsection{Attempts toward N=8 CS theory}
The first attempt to formulate the appropriate CS theory was undertaken by J.~Schwarz in 2004 \cite{Schwarz04}.
He used ${\cal N}=2, 3D$ superfield formalism and tried to construct ${\cal N}=8$ superconformal CS theory as
${\cal N}=2$ CS theory plus {\bf 4} complex matter chiral superfields (with the off-shell content consisting of
{\bf 8} physical bosons, {\bf 16} fermions and {\bf 8} auxiliary fields). However, these attempts failed.
As became clear later, the reason for this failure is that the standard assumption that both matter and
gauge fields are in the adjoint of the gauge group prove to be wrong in this specific case.

Such a theory was constructed by Bagger and Lambert \cite{BLG} and Gustavsson \cite{Gust}. The  basic assumption
of BLG was that the scalar fields and fermions take values in an unusual ``three-algebra''
\be
[T_a, T_b, T_c] = f_{abc}^{\;\;\;\;\;\; d}\,T_d\,. \lb{3Al}
\ee
The gauge group acts as automorphisms of this algebra, gauge fields being still in the adjoint.
The totally antisymmetric ``structure'' constants of the 3-algebra should satisfy a fundamental Jacobi-type identity
\be
f_{abc}^{\;\;\;\;\; d}\,f^{egh}_{\;\;\;\;\;\; d} + \mbox{some permutations of indices} = 0\,. \lb{Ident}
\ee

BLG managed to define ${\cal N}=8$ (on-shell) supersymmetry in such a system and to construct the invariant Lagrangian
\bea
{\cal L}_{N=8} = \tilde{\cal L}_{cs}(A) + \mbox{covariantized kin.terms of $\phi^I, \psi^A$}
+ \mbox{6-th order potential of $\phi^I$} + ...\,, \nonumber
\eea
where $\tilde{\cal L}_{cs}(A)$ is some generalization of the Lagrangian in \p{CS}. All terms involve the constants $f_{abc}^{\;\;\;\;\;\; d}$
and contain only one free parameter, the CS level $k$.

\subsection{Problems with the BLG construction}
Assuming that the 3-algebra is finite-dimensional and no ghosts are present among the scalar fields, the only solution
of the fundamental identity \p{Ident} proved to be $f^{abcd} = \epsilon^{abcd}, a,b = 1,2,3,4$.
\vspace{0.1cm}

Thus the only admissible gauge group is $SO(4) \sim SU(2)_L\times SU(2)_R$ and $\phi^I, \psi^A$ are in the ``bi-fundamental''
representation of this gauge group (in fact these are just  $SO(4)$ vectors). No generalization to the higher-dimensional
gauge groups with the finite number of generators and positive-defined Killing metric is possible.
\vspace{0.1cm}

The $SU(2)\times SU(2)$ gauge group case can be shown to correspond just to two M2- branes. How to describe
the system of $N$ M2-branes?

\subsection{Way out: ABJM construction}
Aharony, Bergman, Jafferis, Maldacena in 2008 \cite{ABJM} proposed a way to evade this restriction on the gauge group.
Their main observation was that there is no need in exotic 3-algebras  to achieve this at all! The fields $\phi^I, \psi^A$
should be always in the bi-fundamental of the gauge group $U(N)\times U(N)$, while the double set of gauge fields should be in the adjoint.

The ABJM theory is in fact dual to M-theory on AdS$_4\times S^7/\mathbb{Z}_k\,$,
and in general it respects only ${\cal N}=6$ supersymmetry and $SO(6)$ R-symmetry. The invariant action is a low-energy limit
of the worldvolume action of $N$ coincident M2-branes on this manifold.

For the gauge group $SU(2)\times SU(2)$, the ABJM theory is equivalent to the BLG theory.

The full on-shell symmetry of the ABJM action is the ${\cal N}=6, 3D$ superconformal symmetry $OSp(6|4)$.
Characteristic features of this action are the presence of sextic scalar potential of special form and the absence of any free parameter
except for the CS level $k$. This $k$ is common for both $U(N)$ CS
actions which should appear with the relative sign minus (only in this case there is an invariance under ${\cal N}=6$ supersymmetry).

\setcounter{equation}{0}

\subsection{Superfield formulations}
Off-shell superfield formulations make manifest underlying supersymmetries and frequently reveal unusual geometric properties of supersymmetric
theories. Thus it was advantageous to find a superfield formulation of the ABJM model with the maximal number
of supersymmetries being manifest and off-shell.

${\cal N}=1$ and ${\cal N}=2$ off-shell superfield formulations
were given in refs. \cite{AdS/CFT} -\cite{Cherkis}.  They allowed
one to partly clarify the origin of the interaction of scalar and
spinor component fields. On-shell ${\cal N}=6$ and ${\cal N}=8$ formulations were also constructed
for both the ABJM and BLG models (see e.g. \cite{Cederwall} - \cite{Lyon}).

The maximally possible off-shell supersymmetry for the CS theory coupled to matter is
${\cal N}=3, 3D$ supersymmetry \cite{BZN3}, \cite{KL}. Thus it was an urgent problem to
reformulate the general ABJM models in ${\cal N}=3, 3D$ superspace. This was recently done
in \cite{BILPSZ}.

This formulation uses the ${\cal N}=3, 3D$ version \cite{BZN3}
of the ${\cal N}=2, 4D$ harmonic superspace \cite{GIKOS}, \cite{book}.

\setcounter{equation}{0}

\section{${\cal N}=3$ superfield formulation of the ABJM model}
\subsection{${\cal N}=3, 3D$ harmonic superspace}
${\cal N}=3, 3D$ harmonic superspace (HSS) is an extension of the standard real ${\cal N}=3, 3D$ superspace by the harmonic variables
parametrizing the sphere $S^2 \sim SU(2)_R/U(1)_R$:
\be
(x^m, \, \theta^{(ik)}_\alpha) \;\Rightarrow \; (x^m, \,\theta^{(ik)}_\alpha, \, u^{\pm}_j)\,, \;\;
u^\pm_i \in SU(2)_R/U(1)_R\,, \; u^{+i}u^-_i = 1\,, \lb{HSS}
\ee
$$
m,n = 0,1,2; \quad i,k,j = 1,2; \quad \alpha =1,2\,.
$$
The most important feature of the ${\cal N}=3, 3D$ HSS is the presence of an analytic subspace in it, with a lesser number of
Grassmann variables (two $3D$ spinors  as opposed to  three  such spinor coordinates of the standard superspace)
\be
(\zeta^M) \equiv (x^m_A, \theta^{++}_\alpha, \theta^0_\alpha, u^\pm_k)\,, \quad \theta^{++}_\alpha = \theta^{(ik)}_\alpha u^+_iu^+_k\,, \;
\theta^{0}_\alpha = \theta^{(ik)}_\alpha u^+_iu^-_k\,. \lb{ASS}
\ee
It is closed under both the ${\cal N}=3, 3D$ Poincar\'e supersymmetry and its superconformal extension $OSp(3|4)$.

All the basic objects of the ${\cal N}=3$ superspace formulation live as unconstrained superfields on this subspace:
\begin{enumerate}

\item Gauge superfields
\be
V^{++}(\zeta), \quad \delta V^{++} = - {\cal D}^{++}\Lambda(\zeta) - [V^{++}, \Lambda]\,, \quad \Lambda = \Lambda(\zeta)\,. \lb{V}
\ee

\item Matter superfields (hypermultiplets)
\be
(q^{+}(\zeta), \;\bar{q}^+(\zeta)), \; q^+ = u^+_if^i + (\theta^{++ \alpha}u^-_k -\theta^{0 \alpha}u^+_k)\psi^k_\alpha
+ \mbox{$\infty$ of aux. fields}\,. \lb{q}
\ee
\end{enumerate}

In eq. \p{V}, ${\cal D}^{++}$ is the analyticity-preserving derivative on the harmonic sphere $S^2$.

\subsection{${\cal N}=3$ action}
The ${\cal N}=3$ superspace formulation of the $U(N)\times U(N)$ ABJM model \cite{BILPSZ} involves:

1. The gauge superfields $V^{++}_L$ and $V^{++}_R$ for the left and right gauge $U(N)$ groups.
Both of them have the following field contents in the Wess-Zumino gauge:
\be
V^{++} \sim \left(A_m, \phi^{(kl)}, \lambda_\alpha, \chi^{(kl)}_\alpha, X^{(kl)} \right), \lb{Cont}
\ee
i.e. ({\bf 8 + 8}) fields.

2. The hypermultiplets  $(q^{+ a})^{\underline{B}}_A\,, \; (\bar{q}^{+ a})_{\underline{B}}^A, a=1,2$,
in the bi-fundamental of $U(N)\times U(N)$: $A = 1, \ldots N; \underline{B} = 1, \ldots N$.
Each hyper $q^{+ a}$ contributes ({\bf 8 + 16}) physical fields off shell (({\bf 8 + 8}) on shell).

The full superfield action is as follows:
\be
S_{N3} = S_{CS}(V^{++}_L) - S_{CS}(V^{++}_R) + \int d\zeta^{(-4)} \,\bar q^+_a \nabla^{++} q^{+ a}\,, \lb{Fact}
\ee
$$
\nabla^{++} q^{+a} = {\cal D}^{++}q^{+ a} + V^{++}_L q^{+ a} - q^{+ a} V^{++}_R\,.
$$

\subsection{Some salient features of the ${\cal N}=3$ formulation}
\begin{itemize}
\item Though the gauge superfield CS actions are given by integrals over the harmonic superspace, their
variations with respect to $V^{++}_L, V^{++}_R$ are represented by integrals over the analytic subspace
\be
\delta S_{CS} = -\frac{ik}{4\pi} \mbox{Tr} \int d\zeta^{(-4)} \delta V^{++} W^{++}\,, \; W^{++} = W^{++}(\zeta), \;
\nabla^{++}W^{++} = 0\,. \lb{Delta}
\ee
As a result, the equations of motion are written solely in terms of analytic superfields in the simple form:
\be
W^{++}_L = -i\frac{4\pi}{k} q^{+ a} \bar{q}^+_a\,, \; W^{++}_R = -i\frac{4\pi}{k} \bar{q}^{+}_a q^{+a}\,, \;
\nabla^{++}q^{+a} = \nabla^{++}\bar{q}^+_a = 0\,.\lb{Eqs}
\ee
\item
The ${\cal N}=3$ superfield action, in contrast to the ${\cal N}=0$, ${\cal N}=1$
and ${\cal N}=2$ superfield  ABJM actions, {\it does not involve} any explicit superfield potential,
only minimal couplings
to the gauge superfields. The correct 6-th order scalar potential emerges {\it on-shell} after eliminating appropriate auxiliary
fields from both the CS and hypermultiplet sectors.
\item
Three hidden supersymmetries completing the manifest ${\cal N}=3$ supersymmetry to ${\cal N}=6$ are realized by
simple transformations
\be
\delta V^{++}_L = \frac{8\pi}{k}\epsilon^{\alpha(ab)}\theta^0_\alpha q^{+}_{ a} \bar{q}^+_b, \,
\delta V^{++}_R = \frac{8\pi}{k}\epsilon^{\alpha(ab)}\theta^0_\alpha \bar{q}^{+}_{ a} \bar{q}^+_b, \,
\delta q^{+ a} = i\epsilon^{\alpha(ab)}\nabla^0_\alpha q^+_b\,, \lb{N6susy}
\ee
where $\nabla^0_\alpha$ is the properly covariantized derivative with respect to $\theta^0_\alpha$.
\item
The hidden R-symmetry transformations extending the R-symmetry of the ${\cal N}=3$ supersymmetry to $SO(6)$ also
have a very transparent representation in terms of the basic analytic superfields.
\item
The ${\cal N}=3$ harmonic superspace formulation makes manifest that the hidden ${\cal N}=6$ supersymmetry is compatible
with other product gauge groups, e.g. with $U(N)\times U(M), N \neq M$, and with other types of the bi-fundamental representation
for the hypermultiplets.
The hidden supersymmetry transformations have the universal form in all cases and suggest a simple criterion as to which gauge groups
admit this hidden supersymmetry. In this way one can e.g. reproduce, at the ${\cal N}=3$ superfield level, the classification
of admissible gauge groups worked out at the component level by Schnabl and Tachikawa in \cite{schn}.
\item
The enhancement of the hidden ${\cal N}=6$ supersymmetry to ${\cal N}=8$ and R-symmetry $SO(6)$ to
$SO(8)$ in the case of the gauge group $SU(2)_k\times SU(2)_{-k}$ is also very easily seen in the
${\cal N}=3$ superfield formulation.
Actually, this enhancement arises already in the case of the gauge group
$U(1)\times U(1)$ with a doubled set of hypermultiplets (with {\bf 16} physical bosons as compared to
{\bf 8} such bosons in the ``minimal''   $U(1)\times U(1)$ case \cite{sch2}).
\end{itemize}

\setcounter{equation}{0}

\section{Outlook}

In conclusion, let me list some further problems which can be studied within the ${\cal N}=3$ superfield formulation sketched above.

\begin{itemize}
\item Construction and study of the quantum effective action of the ABJM-type models in the ${\cal N}=3$
superfield formulation. The fact that the superfield equations of motion are given solely in the analytic subspace hopefully implies
some powerful non-renormalizability theorems \cite{QuantN3}.
\vspace{0.2cm}

\item Computing the correlation functions of composite operators directly in the ${\cal N}=3$
superfield approach as comprehensive checks of the considered version of the AdS$_4/CFT_3$ correspondence.
\vspace{0.2cm}

\item A study of interrelations between the low-energy actions of M2- and D2-branes using the Higgs mechanism \cite{Higgs1},
in which the second system is interpreted as a Higgs phase of the first one.
\vspace{0.2cm}

\item Constructing the full effective actions of M2-branes in terms of the ${\cal N}=3$ superfields (with a Nambu-Goto action
for scalar fields in the case of one M2-brane and its nonabelian generalization for $N$ branes).
\vspace{0.2cm}

\item ETC ...

\end{itemize}

\section*{Acknowledgements}

\noindent I thank the Organizers of Jiri Niederle's Fest for inviting me to present this talk and my co-authors in refs. \cite{BILPSZ}
and \cite{QuantN3} for our fruitful collaboration.
I acknowledge a support from grants of the Votruba-Blokhintsev and the Heisenberg-Landau Programs, as well as from the RFBR
grants 08-02-90490, 09-02-01209 and 09-01-93107.

\end{document}